\documentclass[12pt]{article}
\usepackage{amsmath,amssymb,amscd,latexsym,euscript,mathrsfs,textcomp}
\usepackage[all]{xy}
\usepackage[cp1251]{inputenc}

\newtheorem{theorem}{\bf Theorem}[section]

\newtheorem{proposition}[theorem]{\bf Proposition}

\newtheorem{definition}{\sc Definition}[section]
\newtheorem{example}[definition]{\sc Example}

\def\leq{\leqslant}

\begin{document}


\begin{center}
 {\Large Time Estimation Model of Concurrent Computing Systems}\footnote{The paper is
 performed under the program of strategic development of state educational
 institution of higher professional education, \textnumero 2011-PR-054}
\end{center}
\centerline{A. A. Husainov, husainov51@yandex.ru}
\centerline{E. S. Kudryashova, ekatt@inbox.ru}

\begin{abstract}
We consider an asynchronous system with transitions corresponding to the instructions
 of a computer system. For each instruction, a runtime is given. We propose a mathematical
 model, allowing us to construct an algorithm for finding the minimum time of the parallel
 process with a given trace. We consider a problem of constructing a parallel process
 which transforms the initial state to given and has the minimum execution time.
 We show that it is reduced to the problem of finding the shortest path in a directed graph 
with edge lengths equal to 1.
\end{abstract}


2000 Mathematics Subject Classification 68Q10, 68Q85

\section*{Introduction}

We study a computing system that have memory locations containing some data and a set of operations (instructions, machine commands), which change the states of the memory. 
Some instructions can be executed concurrently. It is known the system state at the initial time. Runtime is defined for each operation.

A sequential process is a sequence of instructions. Our first task is to find this process parallelization algorithm, which would have calculated instructions for each time point. The second task is to specify the minimum time to reach a given state of memory from the initial state.

\section{ Basic notations and definitions}

{\em Asynchronous system} \cite{1} is a quintuple
$A = (S, s_0, E, I ,Tran)$
consisting of the set $S$ of {\em states},  {\em initial state} $s_0\in S$, set of the {\em instructions} $E$, subset
$Tran\subseteq S\times E\times S$ of {\em transitions}, 
and the irreflexive symmetric
{\em independence} relation $I\subseteq E\times E$ which satisfy to conditions

1. If $(s,a,s')\in Tran ~\&~ (s,a, s'')\in Tran$ then $s'=s'' $.

2. For all $s\in S$, if $(a,b)\in I ~\&~ (s,a,s')\in Tran ~\&~ (s',b,s'')\in Tran$, then
there is $s_1\in S$ such that $(s,b,s_1)\in Tran ~\&~ (s_1,a,s'')\in Tran$.

In particular, each Petri net can be considered as an asynchronous system whose states are markings and instructions are transitions. Independence relation consists of the pairs of transitions that have no common places.

Let $E$ be a set and let $I\subseteq E\times E$ be an irreflexive symmetric relation. The elements $a,b\in E$ are {\em independent} if $(a,b)\in I$. The equivalence relation consisting of pairs of words produced from each other by using a series of permutations of adjacent independent letters is defined on the monoid of words $E^*$.
{\em Trace} is the equivalence class $[w]$ for a word $w \in E^*$.
It is easy to see that the operation on the traces defined by the rule $[w_1][w_2]=[w_1w_2]$ turns the set of equivalence classes in the monoid. This monoid is denoted by $M(E,I)$ and is called {\em trace monoid} or a {\em free partially commutative monoid}.

Traces $[w_1]$, $[w_2] \in M(E,I)$ are called {\em parallel} if for any letter $a_1$ of the word $w_1$ and $a_2$ of $w_2$ we have $(a_1,a_2) \in I$. It is known \cite{2} that any asynchronous system $A = (S, s_0, E, I ,Tran)$ can be defined as a set $S$ with a partial right action of a monoid $M(E, I)$. The action is given by $s\cdot a = s'$ if $(s,a,s') \in Tran$. Action $s\cdot a$ undefined if there is no $s'$ satisfies the condition $(s,a,s')\in Tran$.

This allows us to consider a morphism of asynchronous systems as a morphism of the corresponding sets with a partial trace monoid action.

\begin{definition}
Homomorphism of asynchronous systems $(\sigma,f): A\to A'$ is a pair consisting of a map $\sigma: S\to S'$ and a homomorphism of monoids $f: M(E,I)\to M(E',I')$ 
satisfying the conditions

1. $f$ maps parallel traces in parallel;

2. $\sigma (s_0)=s_0'$;

3. $\sigma (s\cdot a)= \sigma (s)\cdot f(a)$ if action $s\cdot a$ is defined.
\end{definition}

Let $A = (S, s_0, E, I, Tran)$ is asynchronous system. 
A {\em function of time} on $A$ is an arbitrary function $\tau: E\to N$ taking values in the set of integers $N = \{0, 1, 2, ...\}$.

Triples $(s,e,s')\in Tran$ are denoted by arrows $s\stackrel{e}\to s'$.
Every sequence of instructions 
$$
s\stackrel{e_1}\to s_1\stackrel{e_2}\to s_2\to ...\to s_{n-1}\stackrel{e_n}\to s_n=s'
$$
consisting of triples belonging $Tran$ we call the {\em process} or {\em path} connecting the states $s$ and $s'$. In this case, the action of the monoid $M(E,I)$
 on $S$ assigns the pair $(s, [e_1...e_n])$ into element $s' \in S$.

\section{The minimum trace runtime}

If the execution times for instructions are the same and equal to 1, 
then the minimum execution time
 for the trace will be equal to the height of its Foata normal form \cite{3}. In general, if the time $\tau(e)\in N$ corresponds to instruction $e\in E$, then we  
 decompose each instruction into the composition of small pairwise dependent instructions 
the execution of which are 1 and apply the algorithm to construct the Foata normal form for the resulting trace.  
These small instructions can be described as an instruction in the expansion which they participate, and the instruction will be equal to $e^{\tau(e)}$. Also we have to enter the intermediate states. For this purpose we introduce a new asynchronous system associated with a function of time.

Let $A$ is asynchronous system with the function $\tau: E\to N$. We define a total order 
relation on the set $E$ and consider an asynchronous system 
$A_\tau = (S_\tau, s_0, E, I, Tran_\tau)$ defined as follows. One has a set of states
\begin{multline*}
S_\tau=\{(s,a_{1}^{i_1},a_{2}^{i_2}...a_{m}^{i_m})~| \\
 s\in S,~
s\cdot a_1 a_2\ldots a_m\in S,~a_1<a_2<...<a_m,~
(a_i,a_j)\in I,\\
~for ~all ~ 1\leq i<j\leq m, ~1\leq i_1<\tau(a_1), \ldots,~1\leq i_m<\tau(a_m)\}.
\end{multline*}
For technical reasons, it will be convenient to consider the states 
$(s,a_{1}^{i_1}a_{2}^{i_2}...a_{m}^{i_m})$ where for some 
$q\in \{1, 2, ..., m\}$ we have $i_q = 0$ or $i_q = \tau (a_q)$. 
They will be identified with the elements of $S_\tau$ using formulas
\begin{gather*}
(s,a_{1}^{i_1}a_{2}^{i_2}...a_{q-1}^{i_{q-1}}a_{q}^{0}a_{q+1}^{i_{q+1}}...a_{m}^{i_m})=
(s,a_{1}^{i_1}a_{2}^{i_2}...a_{q-1}^{i_{q-1}}a_{q+1}^{i_{q+1}}...a_{m}^{i_m})\\
(s,a_{1}^{i_1}a_{2}^{i_2}...a_{q-1}^{i_{q-1}}a_{q}^{\tau(a_q)}a_{q+1}^{i_{q+1}}...a_{m}^{i_m})=
(s\cdot a_q,a_{1}^{i_1}a_{2}^{i_2}...a_{q-1}^{i_{q-1}}a_{q+1}^{i_{q+1}}...a_{m}^{i_m}).
\end{gather*}

We define a partial action of the monoid $M(E,I)$ on the $S_\tau$ considering
$$(s,a_{1}^{i_1}a_{2}^{i_2}...a_{m}^{i_m})\cdot a =
(s,a_{1}^{i_1}a_{2}^{i_2}...a_{q-1}^{i_{q-1}}a_{q}^{i_{q}}a_{q+1}^{i_{q+1}}...a_{m}^{i_m})$$
if $a=a_q$ for some $q\in \{1,2,...,m\}$. If $(a,a_r)\in I$ for all $r\in \{1,2,...,m\}$,
 then we insert an element $a\in E$ in sequence so that there are inequalities $a_1<a_2<...<a_{q-1}<a<a_q<...<a_m$ for some $q$ and let $(s,a_{1}^{i_1}a_{2}^{i_2}...a_{m}^{i_m})\cdot a =
(s,a_{1}^{i_1}a_{2}^{i_2}...a_{q-1}^{i_{q-1}}aa_{q}^{i_{q}}...a_{m}^{i_m})$.

Action is not defined in the other cases.

Define the sets mapping  $i: S\to S_\tau$ by the formula $i(s) = (s, 1)$. Let $t: M(E,I)\to M(E,I)$ is homomorphism which is defined by values on the elements $a\in E$ equal to $t(a) = a^{\tau(a)}$.

\begin{proposition}
Pair $(i,t)$ is homomorphism of the asynchronous systems $A\to A_\tau$.
\end{proposition}

{\em Parallel process, which realizes the trace} $\mu $, is the composition of traces
$$
[a_{i_1}a_{i_2}...a_{i_p}][a_{j_1}a_{j_2}...a_{j_p}]...[a_{k_1}a_{k_2}...a_{k_r}]=\mu
$$
which is equal to the trace and consisting of units within each of which the directions pairwise independent.

\begin{proposition}
The minimum runtime of trace $[a_1a_2...a_n]$, transforming the system from state $s$ to a state $s'$, is equal to the height of the Foata normal form of trace $[a_1^{\tau (a_1)}a_2^{\tau (a_2)}...a_n^{\tau (a_n)}]$. Parallel process with minimum time is equal to that normal form.
\end{proposition}

\begin{example}
Let us consider the pipeline Petri net consisting of three operating units
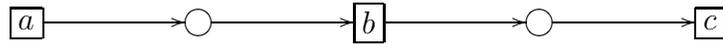
\begin{figure}[h]
$$
\xymatrix{
&*+[F]{a} \ar[rr]&&*+<10pt>[o][F]{ } \ar[rr]&&*+[F]{b} \ar[rr]&&*+<10pt>[o][F]{ } \ar[rr]&&*+[F]{c}}
$$
\caption{The pipeline Petri net}\label{pic1}
\end{figure}
\end{example}

Let the execution times are $\tau(a)=3,\tau(b)=1,\tau(c)=2$. If the input received $n$ numbers the time process trace will be $[(a^3bc^2)^n]$. It is easy to see that the Foata normal form is
$$[a][a][a]([b][ac][ac]a])^{n-1}[b][c][c]$$
Its height is equal to $4n+2$. Hence the minimum runtime using three processors power is $T_3=4n+2$. Runtime on a single processor is $T_1 = 6n$. Consequently, the average acceleration power is $6n/(4n+2)\approx 3/2$.

\section{Search for a parallel process with minimal time 
to achieve a given reachable state from the initial state}

Let us consider an asynchronous system $A$ with the function of time $\tau: E\to N$. 
Let $A_\tau$ is the corresponding asynchronous system. 
We construct a directed graph whose vertex set is $S_\tau$. If
$$(s,a_1^{i_1}a_2^{i_2}...a_p^{i_p})\cdot e_1\cdot e_2...\cdot e_n=(s',b_1^{j_1}b_2^{j_2}...b_q^{j_q})$$
for some vertices $(s,a_1^{i_1}a_2^{i_2}...a_p^{i_p})\in S_\tau$,$ (s',b_1^{j_1}b_2^{j_2}...b_q^{j_q})\in S_\tau$ 
and such $e_1,e_2,...,e_n\in E $ that $(e_i,e_j)\in I $ for all $1\leq i<j\leq n$, 
then these vertices are joined by directional arrow of length 1.

The elements $s\in S$ are identified with pairs $(s,1)\in S_\tau$ 
where 1 is the neutral element of the monoid $M(E,I)$.

\begin{proposition}
A parallel process of the minimum time that takes the system $A$ 
from state $s_0$ to state $s$ corresponds to the shortest path 
in the constructed graph connecting vertices $(s_0, 1)$ and $(s, 1)$.
\end{proposition}

Algorithms for finding the shortest directed path are well known. For example, the vertices are colored with the colors $0,1,2,...$ as follows: first, the vertex $s_0$ is painted the color of 0. Then unpainted ends coming out of her arrows are painted color of 1. Then unpainted ends of arrows coming out of the vertex colors of 1 are painted color of 2, etc. until we color the top $s$. Vertex color $s$ will be the shortest path length. A slight modification of the algorithm leads to a method of finding a path of minimum length.

\section*{Conclusion}
The proposed time model $A_\tau$ can be interpreted 
as a discrete model of  E. Goubault timed automaton \cite{4}. 
A similar model can be constructed for the distributed asynchronous automata entered in \cite{5}. 
But in order to enable it to build algorithms for time estimates, it is necessary
 to involve some additional conditions on these automata.


\begin{thebibliography}{1}

\bibitem{1}
M. Bednarczyk.
\newblock{``Categories of Asynchronous Systems'',}
PhD Thesis, Brighton: University of Sussex, 1987.

\bibitem{2}
A. A. Husainov.
\newblock{``On the homology of small categories and asynchronous transition systems'',}
Homology Homotopy Appl, 2004. V.1, \textnumero 6. P. 439-471.

\bibitem{3}
V. Diekert.
\newblock{``Combinatorics on Traces'',}
Lecture Notes in Computer Science, 454, Spring-er-Verlag, Berlin, 1990.

\bibitem{4}
E. Goubault.
\newblock{``Durations for truly-concurrent transitions'',}
Programming Languages and Systems - ESOP '96,  Lecture Notes in Computer Science,  1058, Springer-Verlag, Berlin, 1996, 173-187.

\bibitem{5}
E. S. Kudryashova, A. A. Khusainov,
\newblock{``Generalized Asynchronous Systems'',}
Modeling and Analysis of Information Systems, 2012.  \textnumero 4. P. 78-86. (Russian)

\end{thebibliography}
\end{document}